\documentclass[seceq]{ptptex}
\newcommand{\D}{\displaystyle}

\usepackage{epsfig}
\usepackage{graphicx}

\title{
Masses, Deformations and Charge Radii---Nuclear Ground-State
Properties in the Relativistic Mean Field Model}
\author{
Lisheng {\sc Geng},$^{1,2,}$\footnote{E-mail:
lsgeng0@rcnp.osaka-u.ac.jp} Hiroshi {\sc
Toki}$^{1,}$\footnote{E-mail: toki@rcnp.osaka-u.ac.jp} and Jie
{\sc Meng}$^{2,}$\footnote{E-mail: mengj@pku.edu.cn} }
\inst{
$^1$Research Center for Nuclear Physics (RCNP), Osaka University,\\
Ibaraki 567-0047, Japan
\\
$^2$School of Physics, Peking University, Beijing 100871, P. R.
China }
 \abst{We perform a systematic study of the ground-state properties of all the nuclei from the proton drip line to the
 neutron drip line throughout the periodic table employing the relativistic mean field model.
 The TMA parameter set is used for the mean-field Lagrangian density, and a state-dependent BCS method is adopted to
 describe the pairing correlation. The ground-state properties of a total of 6969 nuclei
with $Z,N\ge 8$ and $Z\le 100$ from the proton drip line to the
neutron drip line, including the binding energies, the separation
energies, the deformations, and the rms charge radii, are
calculated and compared with existing experimental data and those
of the FRDM and HFB-2 mass formulae. This study provides the first
complete picture of the current status of the descriptions of
nuclear ground-state properties in the relativistic mean field
model. The deviations from existing experimental data indicate
either that new degrees of freedom are needed, such as triaxial
deformations, or that serious effort is needed to improve the
current formulation of the relativistic mean field model.}
\begin{document}
 \maketitle

\section {Introduction}
 The relativistic mean field (RMF) model has achieved great success
 during the last twenty years \cite{rmf}.  Starting from an
effective Lagrangian density, the nucleons are described as
structureless point particles moving independently in the nuclear
potential generated by exchanging various mesons, including the
isoscalar-scalar $\sigma$ meson, the isoscalar-vector $\omega$
meson, the isovector-vector $\rho$ meson and the photon. Using the
classical variational principle, one can obtain the Dirac
equations for nucleons and the Klein-Gordon equations for mesons.
These equations are very difficult to solve, and for this reason,
usually two simplifications have to be made in any practical
application, i.e. the mean-field approximation and the no-sea
approximation. The mean-field approximation treats the meson
fields as classical $c$-numbers, and the no-sea approximation
implies that the vacuum is not polarized. The large attractive
potential provided by the $\sigma$ meson and the large repulsive
potential provided by the $\omega$ meson naturally generate the
spin-orbit potential, and therefore they lead to proper shell
structures for finite nuclei without any additional parameters.
The Dirac structure has more physical significance than it seems
to have. For example, recently, the origin of the pseudospin
symmetry has been naturally explained as a relativistic symmetry
in the RMF model \cite{pseudo1,pseudo2}.

There are very few parameters in the RMF model, and all of them
have clear physical meanings. These parameters include the masses
of the nucleons and the mesons and the coupling constants of the
meson fields, and are usually fitted to the saturation properties
of nuclear matter and the masses and charge radii of a few
selected spherical nuclei. Based on such a simple physical
picture, the RMF model has successfully described numerous nuclear
properties in various regions: from stable nuclei
\cite{gambhir.90} to unstable nuclei
\cite{hirata90,mengprl.96,mengprl.98}; from the very light halo
nucleus $^{11}$Li \cite{mengprl.96} to the latest superheavy
nucleus $^{288}115$ \cite{gengprc.03}; from neutron (proton) skins
\cite{gengnpa.04} to proton emitters \cite{gengptp.04}. Moreover,
the successful descriptions of the anomalous kinks in the isotope
shifts of Kr, Sr and Pb \cite{sharma.93} nuclei demonstrate that
the RMF model can properly reproduce the deformations and the
inherent shell effects of these nuclei, which altogether determine
the anomalous kinks. However, these anomalous kinks are difficult
to explain in the non-relativistic Hartree-Fock models. In
addition to all these works on nuclear ground-state properties,
there are equally many successful applications in other fields of
nuclear physics, including identical bands in super-deformed
nuclei \cite{ring.93}, collective multipole excitations
\cite{ma.02}, hypernuclei\cite{lv.03}, and neutron stars and
supernovae \cite{shen.98}, to name just a few.

Despite all the successes of the RMF model, a systematic study of
all the nuclei throughout the periodic table from the proton drip
line to the neutron drip line is still missing, due to several
difficulties encountered in previous attempts. Firstly, a
self-consistent calculation of the RMF model for deformed nuclei
is very time consuming. Secondly, and more importantly, an
efficient and economical pairing method that can treat all the
nuclei from the proton drip line to the neutron drip line was not
well known until recently. This is the reason that in 1997 Hirata
et al. constructed a mass table for 2174 even-even nuclei with
$8\le Z\le 120$ without including the pairing correlation
\cite{hirata.97}, and then, in 1999, Lalazissis et al. developed a
mass table for 1315 even-even nuclei with $10\le Z\le 98$ adopting
a constant-gap BCS method \cite{lala.99}.

In recent years, a state-dependent BCS method with a zero-range
$\delta$-force has been found to be an efficient and economical
method to treat the pairing correlations for all the nuclei from
the proton drip line to the neutron drip line
\cite{delta1,delta2,gengptp.03}. This method not only can treat
the continuum of neutron-rich nuclei reasonably well by taking
into account the contribution of resonant states that have large
overlaps with the occupied states below the Fermi surface, but
also can avoid the problem of the convergence difficulty usually
encountered in the constant-gap BCS method when moving away from
the line of $\beta$ stability. This method has been introduced
into the axially-deformed RMF+BCS method \cite{gengptp.03}, and it
has been demonstrated to work well in almost the entire mass
region \cite{gengprc.03,gengnpa.04,gengptp.04,gengmpa.04}. With
this method, we now are ready to perform a systematic study of the
ground-state properties of  all the nuclei with $Z,N\ge8$ and
$Z\le 100$ from the proton drip line to the neutron drip line in
the RMF model.

Such a systematic study is urgently needed, because
experimentally, with developments in radioactive nuclear beams
(RNB) \cite{rnb}, more and more exotic nuclei can be investigated,
and it is expected that more exotic phenomena, such as the
so-called giant halos \cite{mengprl.98}, will be discovered.
Therefore, a theoretical model that can make reliable predictions,
can successfully describe various new phenomena, and is also based
on sound physical grounds is badly needed. Theoretically, to
improve the present formulation of the RMF model, to better
understand its advantages and deficiencies, to get to know where
and how to make further improvements, and finally to advance our
understanding of nuclear structure, a systematic study of all the
nuclei from the proton drip line to the neutron drip line within
the RMF framework is needed.

This paper is organized as follows. First, we present the
formulations and the numerical details in \S2. Then, in \S3, we
compare the predictions of our calculations, including the binding
energies, the separation energies, the deformations, and the
charge radii, with existing experimental data, those of the
finite-range droplet model (FRDM), and those of the
Hartree-Fock-Bogoliubov (HFB) mass formula. The work is summarized
in \S4.

\section{Formulations and numerical details}

The relativistic mean field (RMF) model has become a standard
method to study nuclear properties. Therefore, here we only
present the Lagrangian density adopted in the present work, which
includes nonlinear terms for both the $\sigma$ and $\omega$
mesons, and all the symbols have their usual meanings:
\begin{eqnarray}
\mathcal{L} &=& \bar \psi (i\gamma^\mu\partial_\mu -M) \psi \nonumber\\
&&+\,\frac{\D 1}{\D
2}\partial_\mu\sigma\partial^\mu\sigma-\frac{\D 1}{\D
2}m_{\sigma}^{2} \sigma^{2}- \frac{\D 1}{ \D
3}g_{2}\sigma^{3}-\frac{\D 1}{\D
4}g_{3}\sigma^{4}-g_{\sigma}\bar\psi
\sigma \psi\nonumber\\
&&-\frac{\D 1}{\D 4}\Omega_{\mu\nu}\Omega^{\mu\nu}+\frac{\D 1}{\D
2}m_\omega^2\omega_\mu\omega^\mu +\frac{\D 1}{\D
4}g_4(\omega_\mu\omega^\mu)^2-g_{\omega}\bar\psi
\gamma^\mu \psi\omega_\mu\nonumber\\
 &&- \frac{\D 1}{\D 4}{R^a}_{\mu\nu}{R^a}^{\mu\nu} +
 \frac{\D 1}{\D 2}m_{\rho}^{2}
 \rho^a_{\mu}\rho^{a\mu}
     -g_{\rho}\bar\psi\gamma_\mu\tau^a \psi\rho^{\mu a} \nonumber\\
      &&- \frac{\D 1}{\D 4}F_{\mu\nu}F^{\mu\nu} -e \bar\psi
      \gamma_\mu\frac{\D 1-\tau_3}{\D 2}A^\mu
      \psi.
\end{eqnarray}

We use the parameter set TMA \cite{sugahara.94} for the mean-field
Lagrangian density, which has proved to be one of the most
successful modern parameter sets. The parameter values of the
effective force TMA and the nuclear matter properties calculated
with it are tabulated in Tables I and II. Here, it is worthwhile
to note that the coupling constants of TMA are mass dependent.
This originates from two other widely used parameter sets, TM1 and
TM2 \cite{sugahara.94}. In the process of developing a new
parameter set to obtain better agreement for the behavior of the
equation of state at high densities with that predicted by the
relativistic Brueckner-Hartree-Fock theory, Sugahara and Toki
realized that it is difficult to obtain good fits of heavy nuclei
and light nuclei simultaneously using only one parameter set. For
this reason, they developed TM1 for heavy nuclei and TM2 for light
nuclei. Further studies led to the parameter set TMA, aiming at a
consistent description of all the nuclei with one parameter set.
From today's vantage point, the mass dependence of the coupling
constants can be viewed as another form of density dependence.
Recent studies indicate that the density dependence of the
coupling constants, perhaps even the meson masses, may be needed
to obtain a more satisfactory description of the experimental
data, although the exact form of such a density dependence is
presently under intense investigation.
\begin{table}[t]
\setlength{\tabcolsep}{1.0
em}\renewcommand{\arraystretch}{1}\caption{Parameters of the
effective force TMA.}\vspace{0.1cm}
\begin{center}
\begin{tabular}{lcl}
\hline\hline
&&TMA\\
\hline
$M_n$&&938.900 MeV\\
$M_p$&&938.900 MeV\\
$m_\sigma$&&519.151 MeV\\
$m_\omega$&&781.950 MeV\\
$m_\rho$&&768.100 MeV\\
$g_\sigma$&&$10.055+3.050/A^{0.4}$\\
$g_\omega$&&$12.842+3.191/A^{0.4}$\\
$g_\rho$&&$3.800+4.644/A^{0.4}$\\
$g_2$&&$-0.328-27.879/A^{0.4}$\\
$g_3$&&$38.862-184.191/A^{0.4}$\\
$g_4$&&$151.590-378.004/A^{0.4}$\\
 \hline\hline
\end{tabular}
\end{center}
\end{table}
\begin{table}[t]
\setlength{\tabcolsep}{0.8
em}\renewcommand{\arraystretch}{1}\caption{Nuclear matter
properties for the effective force TMA.}\vspace{0.1cm}
\begin{center}
\begin{tabular}{lcccccr}
\hline\hline
&&&&&TMA\\
\hline
$\rho_{\mathrm sat}$ (fm$^{-3}$)&&&&&0.147\\
$E_b$ (MeV)&&&&&$-16.025$\\
$K$ (MeV)&&&&& 318.146\\
$J$ (MeV)&&&&& 30.661\\
$M^*/M(n)$&&&&&0.635\\
$M^*/M(p)$&&&&&0.635\\
 \hline\hline
\end{tabular}
\end{center}
\end{table}

The RMF equations can be solved using the expansion method with
the harmonic-oscillator basis \cite{gambhir.90,gengptp.03}.
Fourteen shells are used for both the fermion fields and the meson
fields, which have been tested and found capable of yielding
reliable convergence in the region in question.

\begin{figure}[tb]
\centering
\includegraphics[scale=0.68]{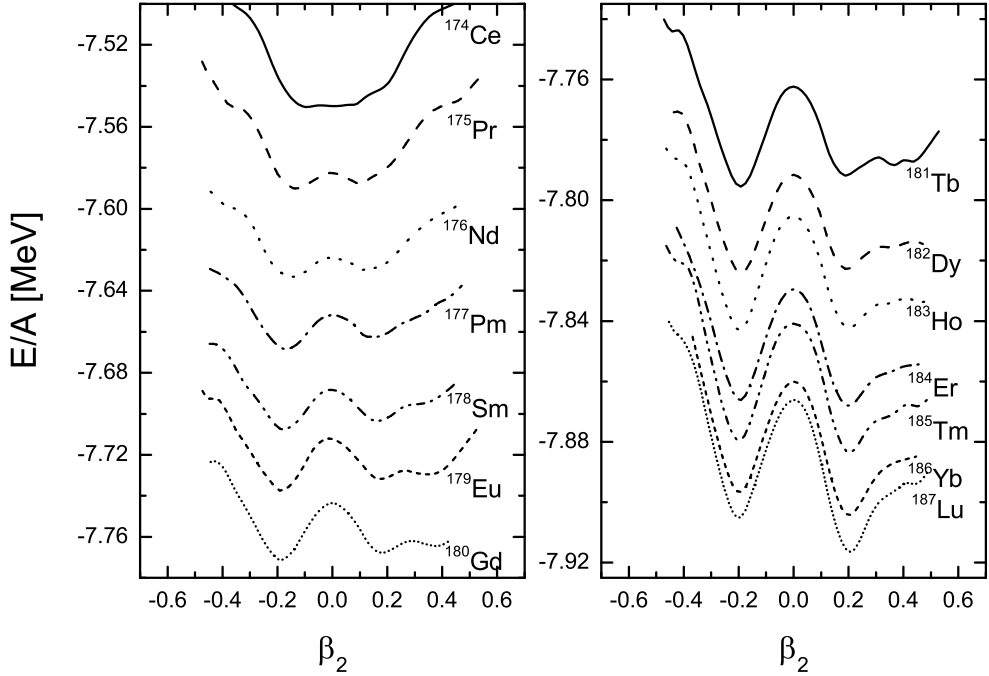}
 \caption{The potential energy surfaces for 14 $N=116$ isotones with $58\le Z\le
 71$, i.e $^{174}_{116}$Ce$_{58}$, $^{175}_{116}$Pr$_{59}$,
 $^{176}_{116}$Nd$_{60}$, $^{177}_{116}$Pm$_{61}$,
 $^{178}_{116}$Sm$_{62}$, $^{179}_{116}$Eu$_{63}$,
 $^{180}_{116}$Gd$_{64}$, $^{181}_{116}$Tb$_{65}$,
 $^{182}_{116}$Dy$_{66}$, $^{183}_{116}$Ho$_{67}$,
 $^{184}_{116}$Er$_{68}$, $^{185}_{116}$Tm$_{69}$,
 $^{186}_{116}$Yb$_{70}$, and $^{187}_{116}$Lu$_{71}$.}
\end{figure}

A zero-range $\delta$-force, $V=-V_0\delta(\vec{r_1}-\vec{r_2})$,
has been used in the particle-particle channel in order to achieve
a consistent and reliable description for all the nuclei from the
proton drip line to the neutron drip line. As usual, in this work
the pairing strength is taken to be the same for both protons and
neutrons, and it is obtained by fitting the experimental odd-even
staggerings with the cutoff $E-\lambda\le 8.0$ MeV, where $E$ is
the single-particle energy and $\lambda$ the chemical potential.
Due to the observed small variation of this quantity with the mass
number $A$ \cite{gengprc.03,gengnpa.04,gengmpa.04}, we have used a
mass-dependent formula for it, $V_0=300+120/A^{1/3}$, in order to
obtain a consistent description for all the nuclei from the
lightest ones to the heaviest ones. Except for the $A^{1/3}$
dependence \cite{satula.98}, there are not many physical reasons
for such a combination of coefficients. Also, we do not exclude
the possibility that using a slightly different formula or
adjusting it more carefully by dividing the entire periodic table
into small areas can improve the final result somewhat. For nuclei
with odd numbers of nucleons, a simple blocking method that does
not break the time-reversal symmetry has been adopted
\cite{gengnpa.04,gengmpa.04,ring.80}.

In order to correctly locate the absolute minimum on the potential
energy surface for each nucleus in the deformation space, a
quadrupole-constrained calculation \cite{flocard.73} has been
performed. We note that this is a rather time-consuming procedure,
but it is nevertheless necessary in order to find the ground-state
configuration with the lowest energy for each nucleus. In Fig. 1,
we display the energy per particle $E/A$ for 14 $N=116$ isotones
with $58\le Z\le 71$ in the rare-earth region as functions of the
deformation parameter $\beta_2$. It is clearly seen how the
deformation evolves as a function of the proton number $Z$: from a
soft spherical shape to oblate, and then to a prolate shape as $Z$
increases from 58 to 71.

We point out here that as a first attempt, neither the mean-field
parameters nor the pairing-channel parameters adopted in this work
have been constrained as stringently as those in most mass tables,
such as FRDM \cite{frdm} and HFB \cite{hfb}. Therefore, the
purpose of this work is not to obtain numbers that agree most
closely with the measured experimental quantities but to obtain a
fairly good description of the experimental data and gain some
useful insight into the entire mass region in order to guide
future works. Theoretically, the deviations from the experimental
data can tell us a lot about the model itself. Some of the small
deviations may be removed with finer adjustments of the
parameters, and others may not. This will elucidate the need of
going beyond the pure mean-field approximation and taking into
account more high-order corrections. For similar reasons, we also
ignore the rotational energy and the Wigner effect, which are more
often treated by adding some phenomenological terms to the final
total energy, except for the center-of-mass correction, which is
approximated by
 \begin{equation}
 E_{\mathrm cm}=-\frac{3}{4}41A^{-\frac{1}{3}}.
 \end{equation}

We have also considered the contribution of the center-of-mass
correction to the root-mean-square (rms) radii of the matter,
proton and neutron density distributions:
 \begin{eqnarray}
 \delta\langle r^2_m\rangle&=&-\frac{\langle
 r^2_m\rangle}{A},\nonumber\\
 \delta\langle r^2_p\rangle&=&-\frac{2\langle
 r^2_p\rangle}{A}+\frac{\langle
 r^2_m\rangle}{A},\nonumber\\
 \delta\langle r^2_n\rangle&=&-\frac{2\langle
 r^2_n\rangle}{A}+\frac{\langle
 r^2_m\rangle}{A}.
 \end{eqnarray}
 Altogether, the corrected rms radii of the matter, proton and
 neutron density distributions are
 \begin{eqnarray}
 R^2_m&\equiv& \langle r^2_m\rangle_\mathrm{corr}=\langle
 r^2_m\rangle+\delta\langle r^2_m\rangle,\nonumber\\
  R^2_p&\equiv& \langle r^2_p\rangle_\mathrm{corr}=\langle
 r^2_p\rangle+\delta\langle r^2_p\rangle,\nonumber\\
  R^2_n&\equiv& \langle r^2_n\rangle_\mathrm{corr}=\langle
 r^2_n\rangle+\delta\langle r^2_n\rangle.
 \end{eqnarray}
 For the rms charge radius, the finite size of the proton is taken into account by folding the point proton density
distribution with a proton charge distribution of Gaussian type.
This leads to
 \begin{equation}
 R^2_c=R^2_p+0.64\quad\mbox{fm}^2.
 \end{equation}
 It should be noted that more effective methods exist
 for treating the finite size of the proton, but because the
 above-described method is that most often used in the literature,
 adopting such a prescription makes it easy to compare the results of our calculations with
 those of existing works. The same is also true for the
 center-of-mass correction to the total energy \cite{bender.00}.

\section{Results and discussion}

\subsection{Nuclear masses}

Making reliable and accurate predictions for nuclear masses over
the entire periodic table has long been a major aim of nuclear
physicists. (Recent reviews of this subject can be found in Ref.
\cite{lunney.03}) Most related nuclear models can be classified
into three categories. The first one consists of macroscopic
models, such as the Bethe-Weizs\"{a}cker mass formula
\cite{heyde.99}. The second one consists of the so-called
macroscopic-microscopic models. One of the best in this category
is the finite-range droplet model (FRDM) \cite{frdm}. The last one
consists of microscopic models, among which the best known are the
Hartree-Fock (HF) method  \cite{hf} and the relativistic mean
field model (RMF) \cite{rmf}. The accuracy of one model is more or
less determined by the number of parameters it employs and the
procedure used to determine parameter values. For any given model,
the more parameters used and/or the more quantities used to
determine the parameters, the more accurate it should be. By the
latter means,  the agreement of the predictions of the
Hartree-Fock method with experimental nuclear masses has recently
been increased to the same level as those of the FRDM mass formula
\cite{hfbcs,hfb}. Comparing the numbers of parameters of different
models is less meaningful, but usually it is believed that the
more microscopic a model, the more reliable it should be when
extrapolated to unknown areas that have not yet been fitted. In
this respect, it is generally accepted that the RMF model is one
of the most promising, due to its explicit Lorentz invariance.

\begin{table}[t]
\setlength{\tabcolsep}{0.4 em}\caption{The rms deviations $\sigma$
between theoretical predictions and experimental data
\cite{audi.03} for nuclear masses. The second column contains the
total number of experimental data considered in the comparison.
The third through the fifth columns contain the rms deviations for
the RMF+BCS model, the FRDM mass formula \cite{frdm}, and the
HFB-2 mass formula \cite{hfb}, respectively. All energies are in
units of MeV.}\vspace{0.1cm}
\begin{center}
\begin{tabular}{cccccc}
\hline\hline
Group&number&$\sigma$(RMF+BCS)&$\sigma$(FRDM)&$\sigma$(HFB-2)\\
 \hline
I&2882&2.118&0.791&0.843\\
II&2157 & 2.108&0.626&0.739\\
III&1960& 2.107&0.617&0.735\\
 \hline\hline
\end{tabular}
\end{center}
\end{table}

One of the quantities often used to describe the overall agreement
of the theoretical predictions and the experimental masses is the
root-mean-square (rms) deviation $\sigma$, defined by
 \begin{equation}\label{eq.1}
 \sigma=\sqrt{\frac{\sum\limits_{i=1}^N \left(M^i_\mathrm{theo}-M^i_\mathrm{exp}\right)^2}{N}},
 \end{equation}
where $N$ is the total number of experimental data. In Table III,
we tabulate this quantity for our calculations, the FRDM mass
formula \cite{frdm}, and the HFB-2 mass formula \cite{hfb}. The
FRDM and HFB-2 mass formulae have been elegantly constructed by
fitting their parameters to masses of more than 1000 known nuclei.
Therefore, their rms deviations $\sigma$ can be considered the
lower limit that any phenomenological model can achieve. (To be
precise, the current lower limit of $\sigma$ for the mass formula
is about 0.3 MeV \cite{lunney.03}.) In order to remove the
influence of the experimental errors, we divide all the 2882
nuclear masses with $Z,N\ge8$ and $Z\le 100$ compiled in Audi 2003
\cite{audi.03} into three groups. The first group, Group I,
includes all the nuclei without any restrictions on their
experimental errors. It contains 2882 nuclei. The second group,
Group II, includes those nuclei whose experimental errors are less
than 0.2 MeV. It contains 2157 nuclei. The last group, Group III,
includes only those nuclei whose experimental errors are less than
0.1 MeV. It contains 1960 nuclei. Here it should be noted that not
all the 2882 nuclear masses are experimental data. Some of them
are extrapolated data using ``systematic trends'' \cite{audi.03}.
However, due to the fact that these data are usually very close to
the experimental data \cite{lunney.03}, we do not make any
distinction here.

Two things can be learned from Table III. First, the FRDM mass
formula, which is the oldest, exhibits the best agreement with the
experiment results, while our calculation has the largest
$\sigma$. The HFB-2 mass formula is similar to the FRDM mass
formula, with its $\sigma$ being about 0.1 MeV larger than that of
the FRDM mass formula. However, noting that both FRDM and HFB-2
used more than 1000 nuclear masses to fit their parameters, it is
fair to say that our results are fairly good. The second point
here is that the $\sigma$ of the FRDM mass formula changes most
when going from the third group to the first group. The reason for
this is not clear, because the TMA parameter set is almost as old
as that of the FRDM mass formula. The $\sigma$ of the HFB-2 mass
formula also changes about 0.1 MeV, compared to 0.01 MeV for our
calculations.

\begin{figure}[t]
\centering
\includegraphics[scale=0.6]{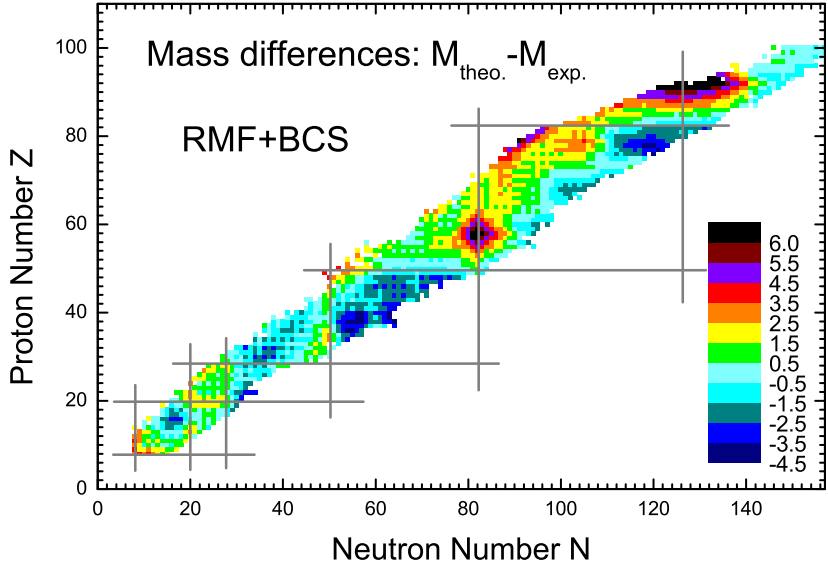}
\caption{Mass differences between the predictions of the present
work and the experimental data for 2157 nuclei whose measured
uncertainties for the masses are less than 0.2 MeV
\cite{audi.03}.}
\end{figure}

To have a much clearer picture of the physics behind $\sigma$, we
plot the mass differences between the results of our calculations
and the experimental data for all the nuclei of Group II in Fig.
2. We have chosen Group II so that the data set is not too small
and the experimental error is not too large to undermine the
analysis. (From this point, we always use the experimental masses
of Group II to make comparisons.) There are several interesting
things that can be learned immediately from this figure, as
discussed below:
\begin{enumerate}
\item Most deviations are in the range of $-2.5$-$2.5$ MeV, which
explains our overall rms deviation of $\sigma\approx2.1$ MeV.

\item There are several strongly overbound regions. The first one
is near $Z=92$ and $N=126$, the second one is near $Z=58$ and
$N=82$, and the third one is near $Z=78$ and $N=92$. Although the
experimental data do exhibit some signs of the magic nature of
these nuclei, apparently, the present RMF model somewhat
overestimates the effect of their magic properties. We stress that
these discrepancies exist not only in the present calculation but
also in all the other modern RMF calculations, regardless of the
parameter sets and other details.

\item In the $Z\ge50$ region, along a certain isotopic chain,
 the proton-rich side is usually overestimated, while the
neutron-rich side is usually underestimated. The two exceptions
are near ($Z=58,N=82$) and in the upper-right corner of the
figure. The behavior of other parameter sets including the latest
ones indicates that this is common to almost all the existing
parameter sets \cite{long.04}.

\item There are also areas where there seems to be strong
underbinding. One is near $Z=38$ and $N=60$, where prolate and
oblate shapes are found to coexist (see Fig. 3). The other region
is near $Z=78$ and $N=120$.

\item More nuclei with $Z\ge50$ seem to be overestimated, while
more nuclei with $28<Z< 50$ seem to be underestimated somewhat.

\end{enumerate}

 We believe that a more extensive study of all these deviations
will clarify whether it is simply a matter of model parameters or
that the pure RMF model itself is too simple to account for all
the underlying physics and therefore that higher-order corrections
need to be introduced. Investigations to clarify this point are in
progress.

One might be tempted to argue that the general underbinding for
nuclei with $28<Z<50$ and the general overbinding for nuclei with
$Z>50$ could be compensated for by adjusting the pairing strength
more carefully. Of course, this may work to some extent. But the
problem is not really that simple, because on the one hand, the
pairing correlations are not large enough to compensate for all
the deviations (particularly the overbinding part), and on the
other hand, the pairing correlations should not be that
complicated. (The most complicated pairing strength ever used in
the zero-range pairing method known to us is that adopted in the
works of Goriely et al. \cite{hfbcs,hfb}, which is slightly
different for protons and neutrons and also depends on whether the
number of nucleons is even or odd.) A more reasonable explanation
is that the shell structure is somewhat incorrect and the
coexistence of prolate and oblate shapes (see discussion below).

There is one more difference between the results of the RMF+BCS,
FRDM and HFB-2 calculations. It is found that although these three
models predict similar results near the line of $\beta$ stability,
their results are quite different for heavy neutron-rich nuclei.
Neutron drip-line nuclei in the RMF+BCS calculation are found to
be more strongly bound (by about 20 MeV) than those in HFB-2,
while those in HFB-2 are also more strongly bound (by about 20
MeV) than those in FRDM. These features should be checked by
experiments with the new generation of radioactive ion beam
facilities.

\subsection{Separation energies}

Compared to the nuclear masses, the one-neutron and two-neutron
separation energies are more important in investigating the
nuclear shell structures and less dependent on the finer
adjustment of the model parameters, because systematic errors can
be cancelled somewhat by subtraction. In Table IV, we tabulate the
rms deviations of the one-neutron separation energies for our
results, for those obtained from the FRDM mass formula, and for
those obtained from the HFB-2 mass formula. One thing that needs
to be noted here is that the experimental error is that of the
one-neutron separation energies, not that of the nuclear masses,
so now the number of nuclei in each group is smaller than that
appearing in Table III. Because the experimental error can more
strongly influence $\sigma$ in this situation, we add another
group (Group IV), which includes only those nuclei for which the
experimental error of the one-neutron separation energies is less
than 0.02 MeV.
\begin{table}[t]
\setlength{\tabcolsep}{0.4 em}\caption{The same as Table III, but
for the one-neutron separation energies.}\vspace{0.1cm}
\begin{center}
\begin{tabular}{cccccc}
\hline\hline
Group&number&$\sigma$(RMF+BCS)&$\sigma$(FRDM)&$\sigma$(HFB-2)\\
 \hline
 I&2790&0.667&0.437&0.509\\
II&1994& 0.640&0.372&0.466\\
III&1767& 0.638&0.369&0.466\\
IV&1030&0.673&0.393&0.506\\
 \hline\hline
\end{tabular}
\end{center}
\end{table}

Most conclusions we derive from the study of the nuclear masses
still hold: The FRDM has the best agreement with experiment, the
HFB-2 is a little worse, and ours still has the largest deviation,
except that now the difference between our calculations and those
of the other two nonrelativistic models has been reduced by about
half. That is to say, for the one-neutron separation energies, our
calculations are similar to those employing the other two mass
formulae. Because in most practical applications, differential
quantities, such as the one-neutron separation energies, are of
more interest than the absolute nuclear masses, this makes our
results comparable to those of the FRDM and HFB-2 mass formulae.

\subsection{Deformations}

The RMF model with modern parameter sets can reproduce the
deformations of finite nuclei very well, particularly those of the
rare-earth nuclei \cite{lala.96}. Experimentally, only for a few
even-even nuclei, deformation parameters can be extracted from the
experimental $BE(2)$ values \cite{raman.01}. However, an overall
$\sigma$ like those used in treating the nuclear masses and the
separation energies will be less useful, because on the one hand,
it is a small quantity with an absolute value less than 1, and on
the other hand, the empirical uncertainty is relatively large.
What are more interesting are the overall general features
exhibited by different models. To see this, we plot the proton
quadrupole deformation parameters for all the nuclei with $Z$,
$N\ge8$ and $Z\le100$ from the three different models, RMF+BCS,
FRDM and HFB-2, in Fig. 3. The following conclusions can be
readily drawn.

\begin{figure}[hp]
\centering
\includegraphics[scale=0.52]{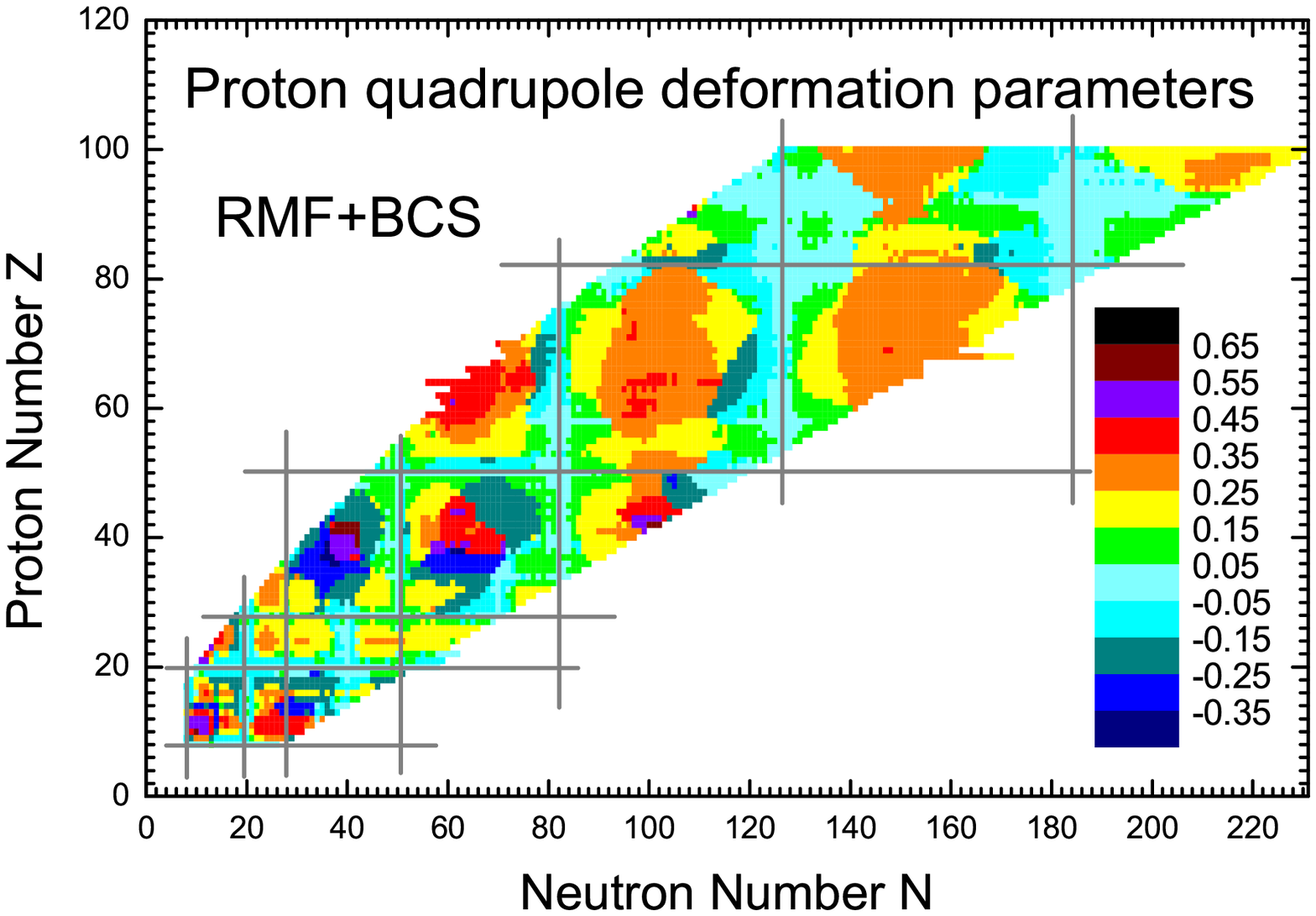}
\includegraphics[scale=0.52]{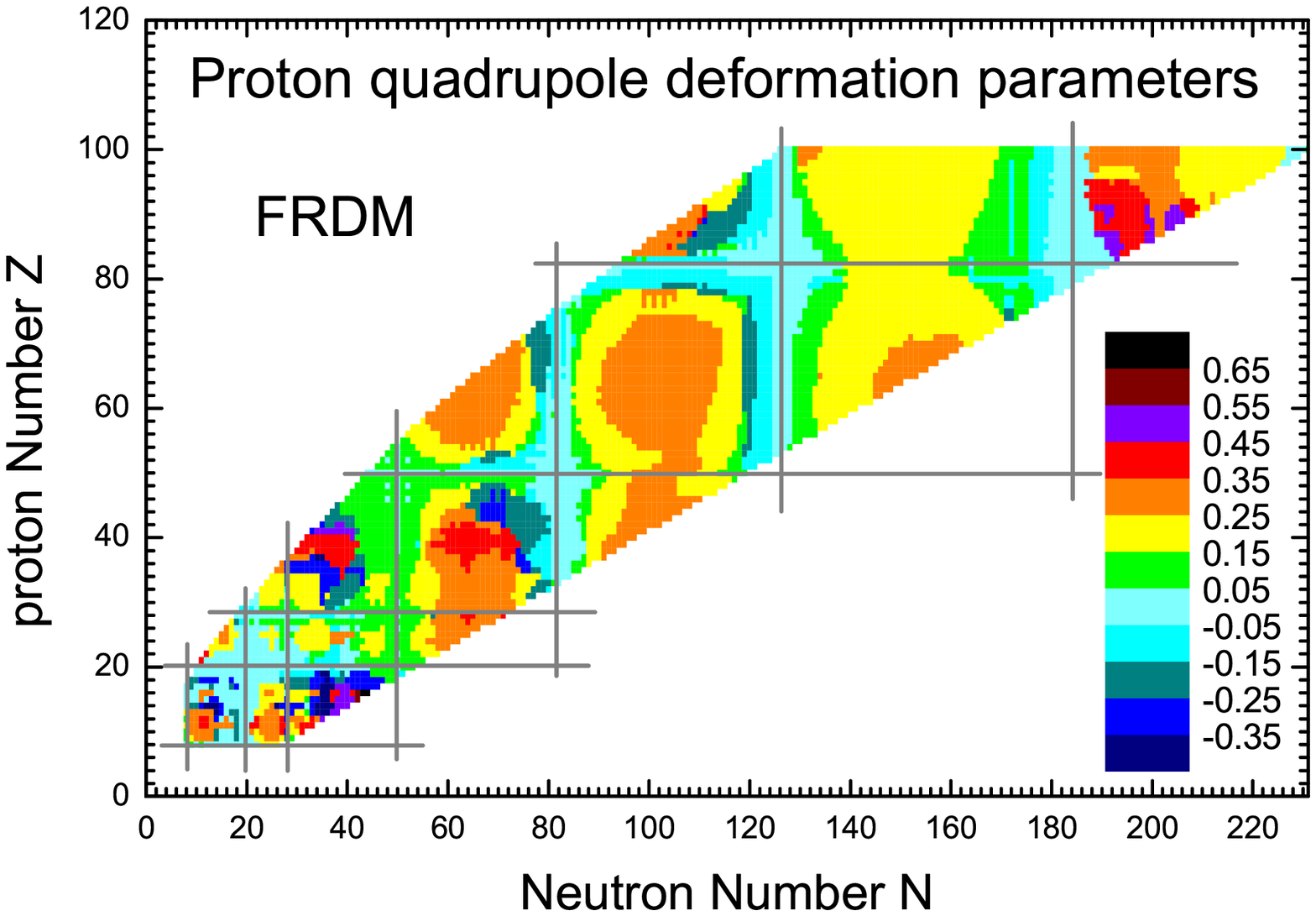}
\includegraphics[scale=0.52]{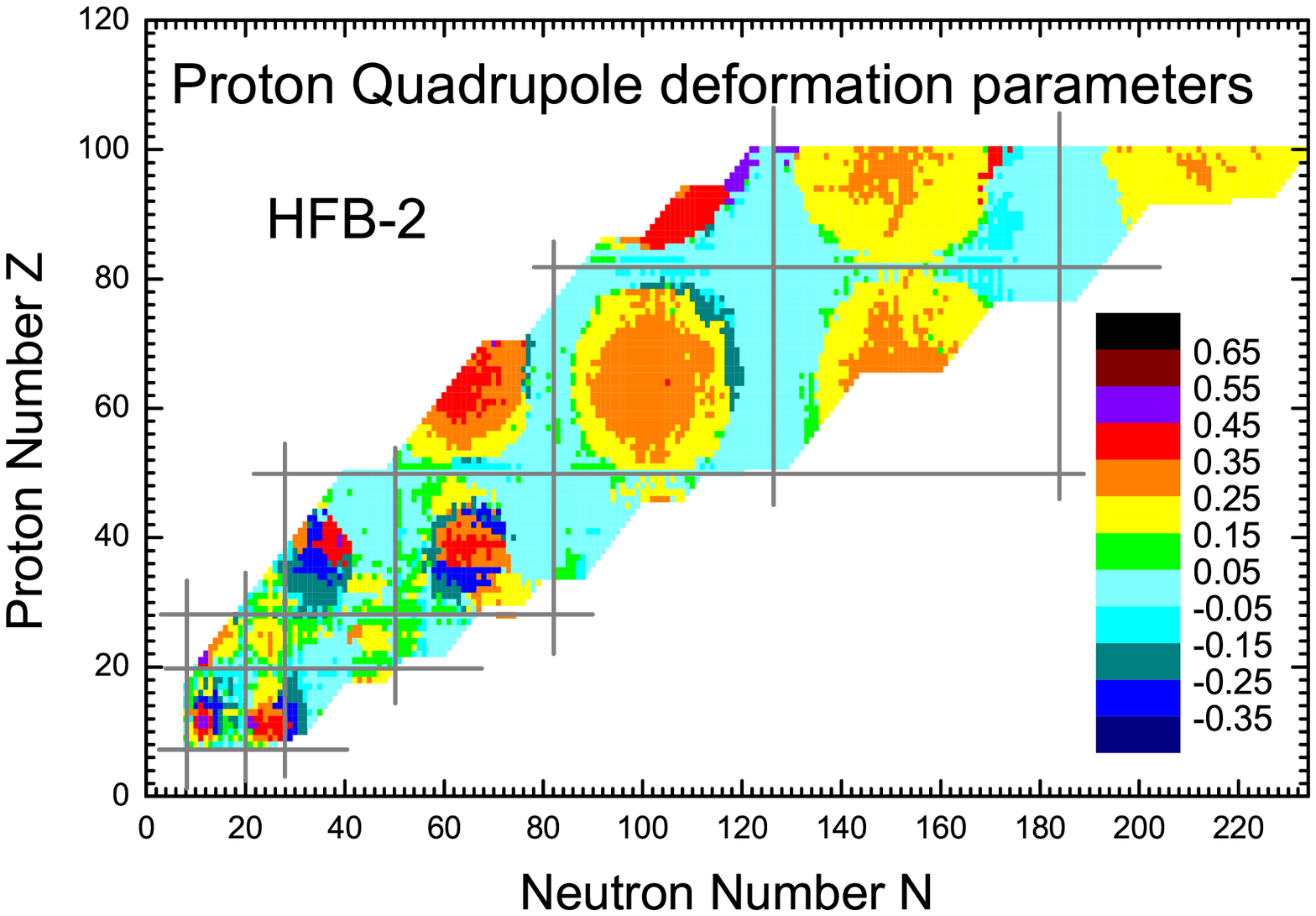}
\caption{The proton quadrupole deformation parameters,
$\beta_{2p}$, for 2157 nuclei whose measured uncertainties for
masses are less than 0.2 MeV \cite{audi.03}. The data obtained in
the RMF+BCS calculations, the FRDM mass formula \cite{frdm}, and
the HFB-2 mass formula \cite{hfb} are displayed in the top,
middle, and bottom panels, respectively.}
\end{figure}

\begin{enumerate}
\item Only very few nuclei are strictly spherical
($-0.05\le\beta_{2p}\le0.05$) and most of them are located at or
near the magic numbers. While RMF+BCS and FRDM exhibit only small
numbers of spherical nuclei, HFB-2 exhibits a large number: More
nuclei near the magic numbers are spherical in the HFB-2 mass
formula than in the other two models. This can be seen quite
clearly from Fig. 3.

\item While isotonic chains with neutron magic numbers
($N=82,126,184$) seem to preserve spherical shapes for the entire
chains, isotopic chains with proton magic numbers are usually
deformed when one moves away from the neutron magic numbers
(except for the $Z=20$ isotopic chain). Among the three models we
employed here, RMF+BCS and FRDM display this feature clearly,
while HFB-2 predicts a large number of spherical nuclei, as
discussed above.

\item Most nuclei with $Z>50$ become prolate deformed when moving
away from the magic numbers either isotonically or isotopically
and have the largest deformations in the middle of major shells.
When approaching the magic neutron number from the lower-$N$ side,
a transition from moderately prolate shapes to moderately oblate
shapes seems to occur often.

\item There are also several regions where strongly prolate and
oblate deformations coexist. The first is near $28<Z<50$ and
$28<N<50$. The second is near $28<Z<50$ and $50<Z<82$. If we
recall what we have found in the study of the nuclear masses, we
note that these regions are where the nuclear masses are a bit
underestimated in our calculations. There seem to be some
connections between these two phenomena: shape coexistence and
underbinding. Since shape coexistence usually implies the
existence of a triaxial deformation, extra degrees of freedom in
the deformation space might be needed to better describe these
nuclei.

\item Another region where prolate and oblate shapes coexist is
near $8<Z<20$ and $8<N<20$. The other is near $8<Z<20$ and
$28<N<50$. The FRDM model exhibits the latter one more clearly.

\item Except for the common features exhibited by all the three
models studied above, there are some features unique to individual
models. One is that FRDM exhibits strong deformation (prolate) in
the upper-right corner of the $(Z=82,N=184)$ intersection, while
the other two models predict more or less spherical shapes in this
region. This difference is interesting for two reasons. First, it
represents a difference between macroscopic-microscopic models and
microscopic models. Second, it influences the magic nature of
$Z=82$ and $N=184$. Another feature unique to HFB-2 is that it
exhibits a strong prolate deformation in the upper-left corner of
the ($Z=82,N=126$) intersection.

\item We found in the above investigation of the nuclear masses
that the regions $(Z\approx92,N\approx126)$ and
$(Z\approx58,N\approx82)$ exhibit the general feature of
overbinding for the case of the RMF model. The same situation is
found here. Looking at Fig. 3, it is easily seen that there is
another isotopic chain, with $Z=92$, exhibiting the feature
(having more spherical nuclei) of conventional proton magic
numbers. The same is also true for $Z=58$, whose magic nature is
especially conspicuous in the two regions near $N=82$ and $N=126$.
\end{enumerate}

To summarize, the general features for the nuclear deformations
exhibited by our calculations are supported by the other two mass
formulae. We also note that there seem to be close connections
between the two nuclear properties, nuclear masses and
deformations. Anomalies appearing in one plot (Fig. 2) can often
be found in the other (Fig. 3) in one way or another. This
provides an interesting way to locate the anomalies and to
understand the underlying physics that causes them. The several
regions in which underbinding exists in connection with shape
coexistence are very good candidates to study the possibility of
triaxial deformations.

\subsection{Charge radii}

In addition to the nuclear deformations, the root-mean-square
(rms) charge radii are also very important quantities to describe
the shapes of finite nuclei. In Table V, we tabulate the rms
deviations of our data from the empirical data compiled by
Nadjakov et al. \cite{nadjakov.94} for 523 nuclei in 42 isotopic
chains. For comparison, the corresponding quantities for the
calculations of the FRDM and HFB-2 mass formulae are also shown.

\begin{table}[htpb]
\setlength{\tabcolsep}{0.8em}\renewcommand{\arraystretch}{0.98}\caption{The
rms deviations $\sigma$ between theoretical predictions and
empirical data \cite{nadjakov.94} for the rms charge radii. The
first column labels the isotopic chains and the second column the
number of isotopes in each chain. Columns 3, 4, and 5 are from the
RMF+BCS calculations, the FRDM mass formula \cite{frdm}, and the
HFB-2 mass formula \cite{hfb}, respectively. The data for the FRDM
mass formula are taken from Ref. \cite{buchi.01}. All radii are in
units of fm.} \vspace{0.1cm}
\begin{center}
\begin{tabular}{ccccc}
\hline\hline
$Z$&number&$\sigma$(RMF+BCS)&$\sigma$(FRDM)&$\sigma$(HFB-2)\\
 \hline
11&12&0.053&0.151&0.056\\
19&10&0.031&0.093&0.011\\
20&10&0.041&0.091&0.023\\
24&4&0.025&0.088&0.006\\
28&5&0.024&0.081&0.010\\
36&8&0.024&0.037&0.027\\
38&22&0.056&0.082&0.025\\
40&5&0.025&0.063&0.010\\
42&7&0.021&0.047&0.029\\
44&7&0.029&0.033&0.036\\
46&6&0.053&0.070&0.019\\
47&6&0.007&0.034&0.020\\
48&17&0.023&0.022&0.042\\
49&24&0.025&0.015&0.038\\
50&18&0.008&0.035&0.015\\
54&20&0.016&0.033&0.015\\
55&29&0.014&0.035&0.015\\
56&28&0.010&0.029&0.016\\
57&3&0.003&0.026&0.004\\
58&4&0.003&0.019&0.014\\
60&16&0.006&0.024&0.017\\
62&17&0.007&0.013&0.025\\
63&22&0.020&0.012&0.031\\
64&8&0.013&0.011&0.017\\
66&18&0.015&0.009&0.033\\
67&14&0.025&0.023&0.013\\
68&12&0.011&0.006&0.026\\
69&17&0.044&0.047&0.033\\
70&18&0.011&0.009&0.022\\
71&4&0.044&0.054&0.058\\
72&7&0.010&0.009&0.006\\
74&5&0.024&0.015&0.009\\
76&7&0.034&0.008&0.017\\
78&14&0.104&0.067&0.034\\
79&15&0.068&0.021&0.025\\
80&26&0.056&0.025&0.031\\
81&18&0.029&0.018&0.012\\
82&22&0.028&0.015&0.019\\
90&5&0.054&0.065&0.048\\
92&5&0.027&0.023&0.024\\
94&6&0.070&0.016&0.021\\
95&2&0.030&0.028&0.021\\
\hline
&523&0.037&0.045&0.028\\
\hline\hline
\end{tabular}
\end{center}
\end{table}

We can learn several things from Table V. First, the overall value
of $\sigma$ for the 523 nuclei in our calculations (0.037 fm) is
just the average of that of the FRDM mass formula (0.045 fm) and
that of the HFB-2 mass formula (0.028 fm). Considering the fact
that our parameters are not constrained as stringently as either
of the other two mass formulae, this result is acceptable, though
further improvements are expected. In fact, our results are even
closer to the empirical data in the rare-earth region than those
of the HFB-2 mass formula. Second, there are several isotopic
chains for which $\sigma$ is larger than 0.05 fm, including
$Z=11$, $Z=38$, $Z=46$, $Z=78$, $Z=79$, $Z=80$, $Z=90$ and $Z=94$.
Recalling our discussion of the nuclear masses and the
deformations, we note that these are either the
magicity-overestimated or prolate-oblate-coexisting regions.

Although not shown here, the much discussed kink in the charge
radii of Pb isotopes can be reproduced very well with our present
calculations, as with almost all the RMF calculations
\cite{long.04}, while all the Hartree-Fock calculations
\cite{hfbcs,hfb}, though very successful in reproducing the
nuclear masses, have failed in this respect.

Another difference between our calculations and those of the HFB-2
mass formula for the nuclear charge radii is that our results are
generally larger than those of HFB-2 by $0.05$-$0.15$ fm for heavy
neutron-rich nuclei with $Z>50$. Not only the absolute charge
radii but also the neutron-skin thickness defined as
$\theta_n=R_n-R_p$, where $R_n$ and $R_p$ are the rms neutron and
proton radii, is found to be somewhat larger than its HF
counterpart. The neutron skin of $^{208}$Pb in our calculations is
0.26 fm, while the HFB-2 mass formula predicts 0.12 fm \cite{hfb}.
Recent experimental results include $0.20\pm0.04$ fm
\cite{star.94}, $0.19\pm0.09$ fm \cite{krasz.99}, and 0.17 fm (no
uncertainties given) \cite{kar.02}. Due to the large experimental
uncertainties, more experimental data are needed before any firm
conclusion can be drawn. It is worthwhile to note that for
$^{208}$Pb, $\theta_n=0.26$ fm and $\theta_n=0.12$ fm are two
typical values in the RMF model and in the HF model
\cite{skin208}. The relation between this quantity and the
density-dependence of the symmetry energy is presently under
active investigation \cite{skin208}. We will study this elsewhere
\cite{symmetry}.

\section{Summary}

We have performed a systematic study of 6969 nuclei with $Z$,
$N\ge 8$ and $Z\le 100$ from the proton drip line to the neutron
drip line in the relativistic mean field model. Comparisons with
two of the most successful existing mass formulae, FRDM and HFB-2,
have been made. Reasonable agreement with the available
experimental data is realized for the nuclear masses, the
separation energies, and the rms charge radii. The general
features of the nuclear deformations were studied together with
those exhibited by the FRDM and HFB-2 mass formulae. Several
regions in which the deviations from the experimental data seem a
bit large have been pointed out, and possible factors causing them
have been discussed.

The overall $\sigma$ of 2.1 MeV for the nuclear masses, though
better than the previous RMF calculation \cite{lala.99} (2.8 MeV),
is still not satisfactory. Based on the present work, further
improvements should be made in order to make the predicted nuclear
masses satisfy the precision required for astrophysical studies.
Possible promising directions include adjusting the parameter
values of the mean-field channel and the pairing channel together,
taking into account the residual corrections ignored in this work,
including the rotational energy and the Wigner effect, and
restoring the broken gauge symmetry of the particle number.
Another possibility is to replace the harmonic-oscillator basis
with the woods-saxon basis in order to make the calculations for
neutron-rich nuclei more reliable \cite{zhou.03}.

A complete calculation that includes all the higher-order
correlations mentioned above is still not very likely in the near
future in the relativistic mean field model. Therefore, our
present study should be seen as a first attempt in this direction
rather than the final work. All future works should first find how
to reduce the overestimated magicities in the several areas that
we discussed. Also, triaxial degrees of freedom might be needed in
the several regions where our study indicates possible shape
coexistence. In this respect, a systematic study of all the nuclei
throughout the periodic table will be very helpful for all future
works along this direction. A more detailed study of the results
of the present calculation is in progress.

\end{document}